# Challenges for Measuring Usefulness of Interactive IR Systems with Log-based Approaches

Daniel Hienert and Peter Mutschke
GESIS – Leibniz Institute for the Social Sciences
Cologne, Germany
{firstname.lastname}@gesis.org

## ABSTRACT
The usefulness evaluation model proposed by Cole et al. in 2009 [2] focuses on the evaluation of interactive IR systems by their support towards the user's overall goal, sub goals and tasks. This is a more human focus of the IR evaluation process than with classical TREC-oriented studies and gives a more holistic view on the IR evaluation process. However, yet there is no formal framework how the usefulness model can be operationalized. Additionally, a lot of information needed for the operationalization is only available in explicit user studies where for example the overall goal and the tasks are prompted from the users or are predefined. Measuring the usefulness of IR systems outside the laboratory is a challenging task as most often only log data of user interaction is available. But, an operationalization of the usefulness model based on interaction data could be applied to diverse systems and evaluation results would be comparable. In this article we discuss the challenges for measuring the usefulness of IIR systems with log-based approaches.

## Categories and Subject Descriptors
• Information retrieval→Users and interactive retrieval

## 1. INTRODUCTION
A rough distinction for the evaluation of Interactive Information Retrieval (IIR) Systems can be made between the human and the system focus. Kelly [8] arranges both perspectives on an IIR evaluation research continuum. System focused studies such as with TREC [11] use a set of topics, documents with relevance judgements where the system's retrieval function is optimized in order to find the most relevant documents. On the other side of the spectrum are user-based evaluations which let the user interact with the system in order to get insight how the system can be improved. There is definitely a big difference between these two kinds of evaluation methods. The first one concentrates on optimizing a system's artefact with the user's intention modelled only in the form of relevance adjustments from assessors. The second one directly takes the user's opinion and the ability to be diverse in experimenting with different systems, functionalities, use cases and situations. The system focused studies have the benefit of having a simple model and measures which can be applied to most systems and are replicable and comparable, but they only focus on the retrieval functionality of the system. The human focused studies with their richness of use cases have the drawback of its costs: user experiments are costly to create, to conduct and to analyze and further are not well applicable to different systems.

Log analysis and log-based evaluation is situated near the middle of Kelly's research continuum and can be a compromise between the system and the human focus. Logs can contain a diversity of signals from the user. For TREC style evaluations implicit feedback from users can be taken as substitute of explicit relevance opinions. For more human focused studies log data can contain information to analyze not only relevance of search results, but also information about the usefulness of other IIR systems components such as recommenders, faceted search or browsing facilities. Additionally, the user's context and task, user actions, navigational paths, search interactions and usage patterns can be extracted. Thereby, the benefit of log-based methods is that they are scalable to different IR systems, different use cases and a lot of users.

In this article we analyze the challenges for measuring the usefulness of IIR systems with log-based approaches. Therefore, we first give a short overview of existing evaluation models and measures in IIR.

## 2. BACKGROUND

### 2.1 Models and Measures for the Evaluation of IIR Systems
The classical IR model used in TREC style studies [11] compares the information need of a user articulated in a search query to results returned by an IR system which is computed by a retrieval function. Ingwersen & Järvelin [6] open up the classical Cranfield paradigm to also include the seeking-, work-, socio-organizational & cultural context which need additional evaluation criteria. Another evaluation model beyond the system view is presented by Tsakonas et Papatheodorou [10] who proposed the triptych framework which combines user-centered, system-centered and content-centered evaluation. Usability then measures the user's interaction with the system, usefulness the user's impression of the content and performance the system's handling with the content. Fuhr [3] proposed a more probabilistic IIR framework where the search process is modelled as transitions between situations and the user has a set of choices in every interaction step. These different evaluation models try to capture various aspects of the IIR process. Measures can be categorized to the following classes according to Kelly [8:99ff]: (1) contextual, (2) interaction, (3) performance and (4) usability.

The *context* is given by the user himself (sex, age, character, personality, mood), his information need or task (type, difficulty, and familiarity), his knowledge (degree, domain expertise, computer or search experience) and the actual situation (work, free time, specific situation, location, time).

The second point describes the *interaction* between the user and the system. Which user actions has been performed, which navigational paths have been gone, which queries issued, which results returned, which documents viewed? This can be described by simple actions and its parameters, but also by more complex search tactics, strategies or with probability models. Interaction data can also be bundled by different user groups, for example by experts vs. novices.

There are a number of measures for measuring *performance* based on interaction data. Classical measures which have been widely used in IR are precision/recall and its extensions. This comes also in combination with the TREC evaluation campaign which provides evaluation data sets with a set of topics and relevance judgements for documents. For IIR these measures have shown a number of limitations: (1) in combination with TREC some measures have been proposed that better fits with the documents found relevant by the user and not the assessor. (2) they only measure the quality of a search result in relation to a single search query. Measuring retrieval quality for higher granularity levels such as the session or the task seems to be more difficult. (3) They do not take into account other artefacts and services of the IR system such as browsing facilities, recommender, system support etc. There are other performance measures such as time-based measures, informativness or utility. *Time-based* measures look at different aspects such as the time spend to complete a search task, to look at a document or to save the first article. *Informativeness* uses a ranked list of documents rated by most informative to least informative. This should give a better distinction than using only binary or graded relevance judgments for single unrelated documents. Another measure is *utility* which uses the subjective user impression of costs (e.g. in $) for different aspects of the search process and system, such as the value of the search result itself or in combination with time spend for getting those results.

The point of *usability* takes more the user's view on the system, described by Kelly [8] as the "evaluative feedback from subjects". Usability can further be divided to the dimensions *effectiveness*, *efficiency* and *satisfaction*. Effectiveness denotes the "accuracy and completeness" of a task and efficiency the "resources expended" for a task [8]. These artefacts of usability may still be derived from simple measures such as time or effort. Other artefacts such as satisfaction, ease of use, ease of learning, usefulness, mental effort, cognitive load, or flow are mostly collected with user questionnaires and are much harder to compute.

As we can see a diversity of measures exist which tries to capture different evaluative artefacts of the interactive IR System. Also there are some models which try to incorporate some of these measures. However, the main challenge, that is still open, is to have a holistic evaluation model which incorporates the magical triangle of user, system and content and which measures the real usefulness of the interactive IR system for the user. Thereby, the points of scalability and comparability are very important. A unique evaluation model based on quantitative information can be used for different systems and let us compare their usefulness on a large scale.

## 2.2 The Usefulness Model

The usefulness evaluation model proposed by Cole et al. [2] tries to evaluate a system by the criterion of "how well the user was able to achieve their goal". This *general goal* derives from a *problematic situation* in which the user is lacking knowledge about a certain topic, experience or situation. The *overall task* is then to seek information which can help to solve the general goal. This general goal can be divided in a sequence of certain *sub goals* which divide the problem in more manageable tasks and therefore *information interactions* the user can process such as collecting information, comparing the results or learning about the topic. Each interaction can further be divided into several *information seeking strategies* (ISS, [1]) which the system side can support such as querying an IR system, receiving results and evaluating the documents. The system's usefulness can then be measured on three levels: (1) the *usefulness of the entire system* towards the overall goal, (2) the *usefulness of each information interaction* towards the sub goal and the overall goal and (3) the *usefulness of the system support* towards each ISS and each goal. In this way the usefulness model tries to measure the system's support towards the overall and each sub goal of the user. This brings together the user's perspective with their problems, goals and sub goals, the resulting interactions and information seeking strategies and the system's perspective with their support towards each ISS (and therefore to each sub goal and the overall goal).

However, in practice the usefulness model has still certain drawbacks for applying it largely for the evaluation of IIR systems without conducting explicit user studies:

(1) In Cole et al.'s usefulness model everything is arranged around the user's initial goal. This is certainly a useful assumption, as it is the systems primary task to optimally support the user in her/his goals. The other way round, from the system's view, it is lacking a lot of knowledge about the user to optimally support her/him. The system has no knowledge about the user's goal, sub goal and tasks, and his context.

(2) There is only initial research how the abstract and theoretical model of usefulness can be operationalized in the form of computational measures. There is no formal framework to derive if an interactive IR system or system components are useful or not. As a starting point Cole et al. [2] formulated a number of research questions at each stage (overall goal, sub goal, ISS) divided by the system's support towards the *results* and *processes* in terms of *correctness*, *effort* and *time*. For example, at the stage of ISS they asked for their example to locate hybrid car information: (1) How useful were suggested terms for formulating queries? [correctness]", (2) "How much were suggested queries/terms used? [effort]". As question 2 can be easily answered, question 1 is much harder to do, as the usefulness can only be measured from the results and in particular from the result's usefulness towards the sub goal and overall goal.

## 3. MEASURING USEFULNESS

Measuring usefulness on the basis of logged interaction data therefore is a challenging task as a lot of information may still be unknown or are hard to capture and to compute from the logs. The purpose of this paper is to discuss these different aspects.

### 3.1 Challenge: Finding the task

The usefulness model compresses the contextual part very much to the artefact of the overall goal, sub goals and tasks. This is certainly the strongest points of the model and makes it simple and effective. Simple, in the sense, that a system does not need to know all the contextual information (sex, age, location, knowledge) but only the goal and sub goals. Effective, in the sense, that everything in the model is derived by the overall goal, such as the tasks, the information interactions, the seeking strategies but also the system's response in supporting these artefacts. However, from the system's point of view it is hard to recognize the overall goal and sub goals, but it seems to be essential. In user studies the overall goal is often predefined in the evaluation task. Additionally, the participant can always be asked how she/he perceived the IIR system's contribution to the overall goal. But with a system focused approach the overall goal and task is mostly missing. Of course, also in running systems the users can be asked for their actual task. However, users often are unable to explain or categorize their actual task or they are drifting from one task to another. A second approach is again trying to extract the task from log data. There has been some research (e.g. [9]) that focuses on constructing the task by user inputs such as query terms. However, the overall impression is that clueing and clustering search queries together is not sufficient to understand the user's task. According to Cole et al. [2] emphasizing that "usefulness … is suited to interaction measurements" we see a much greater but challenging potential to better understand a users' task in analyzing interaction data: How does the user's interactions form a task? and how several tasks may form sub goals and the overall goal. We believe that by taking the perspective of interaction the user's intention can much better be captured than by just looking at query terms.

## 3.2 Challenge: Capturing interaction

Logs are certainly good in measuring interaction data and in fact there has been much research about it. In IIR all activities and processes can be measured which happens between the user and the system such as accessed websites, invoked functionality in the system, and user inputs from keyboard and mouse. With different preprocessing step this basic log data can be transformed to (navigational) paths, user actions with parameters (duration, start-, stop times, search queries) and can be assigned to sessions and preliminarily search tasks.

A prerequisite to make the usefulness model computable is to have a data base which allows measuring the system's support with its results and processes and on the three levels of IIS, sub goal and overall goal. So, what we first need is a framework which captures these signals from arbitrary logs and interprets them. In [4] we proposed a first attempt with the WHOSE toolkit for whole session analysis in IIR. The system was designed to directly load log data from different real world systems. The log data is then preprocessed from unstructured formats to well-defined user actions and its parameters due to a supplementary mapping. Actions are ordered within user sessions and if the user id is known also by certain users. The main outcome for measuring usefulness here is that we can create a structured data set of user interactions out of any IR system's log data. The major challenge we see here is to establish a common standard on logging interaction.

## 3.3 Challenge: Understanding interaction

Once we have a set of structured interaction data, there is the question of interpretation. The main criterion in the usefulness model is to accomplish the overall goal and their sub goals. So, a framework should ideally first extract the goal, sub goal and task from the interaction data and then in a second step compute the success concerning results and processes towards the task. The prominent precision/recall can be one of several measures that help to identify the relevance of search results to the user's task. Besides that, Cole et al. [2] proposed correctness, effort and time concerning processes and results of the IR system at the different levels of the overall goal, sub goal and information seeking strategies. Measuring time and effort is certainly a doable task, but measuring correctness is surely harder to do at the different levels. Cole et al. [2] for example asked "How *useful* were suggested queries/terms for formulating queries on the level of ISS, sub goal and overall goal?". Here, the usefulness of an IIR service such as a query/term recommender for the sub goal of locating information is surely harder to measure.

*Sub-Challenge: Understanding Interaction Data on the Level of System Support*

The usefulness model asked on the third level for the usefulness of system support towards each ISS and each goal. In [5] we did a first attempt to analyze the usefulness of IIR services. We want to measure the effect an IIR service has locally in the moment of usage, but also which effect the service has on the whole search process and therefore on the sub goal. We therefore proposed two variants of usefulness measures, one addressing local and one addressing global effects of a service. Local usefulness is a usage-based measure and describes the usage frequency of the IIR service. Global usefulness is a success-based measure and describes how often positive signals occur in a later phase of the search process after the usage of the IR service, for example bookmarking a record after the usage of a term recommender. Together local and global usefulness can then describe the support given in terms of usage and success towards an information seeking strategy of a user, e.g. supporting the query formulation process with a term recommender. Here, on the level of system support we actually do not need to know the user task because the ISS and sub goal is predefined by the service in question, e.g. the ISS may be "querying" for a term recommender and the sub goal is "locating information about x". The challenging issue here is to define a set of relevant user interactions from log data and to establish a consensus on the set of positive signals to be used as indicators for search success on the level of a single ISS or on the level of the overall system.

*Sub-Challenge: Modelling Signals as Indicators of Search Success*

Interaction data has been used to grasp the user's intention. There has been for example research about what signals to take for implicit feedback as substitutions of relevance adjustments from users (e.g. [7]), others use signals to find navigational problems or defining user preferences. For measuring usefulness on the third level we need signals that describe the user's impression of negative or positive impact of the system's support for a single ISS. These signals can be different depending on the ISS (querying, evaluating docs, savings docs) and depending on the system. Signals can also have several dimensions: the type of user action, its point in time within the task or search process, its strength (weak vs. strong), its frequency and its relevance towards the goal of the ISS. The best way to identify signals and their different aspects for the diverse information seeking strategies are user studies. The user's impression of the system's support has to strongly correlate with these quantitative signals. Then we can use these signals (from logs) as indications of success and non-success of system support. However, to enable the comparison of IIR services in two different systems, the challenging issue is to model the different dimensions of signals in log data with respect to system support.

*Sub-Challenge: Measuring Usefulness on the Levels of Sub Goals and the Overall Goal*

Having a measure of system support on the level of an ISS does not imply having an indicator on the level of support towards the sub goal and overall goal. This has several reasons: (1) again, we first need to know the goal and sub goals. This may be complicated to extract from logs, especially for more complex information needs. (2) It may be hard to measure what's the actual influence of a supporting service towards a sub goal. Thus, the challenging issue is to deconstruct the overall goal to sub goals and sub goals to single ISSs, and, the other way around, to extrapolate the usefulness of the system for achieving sub goals and, finally, the entire information seeking task from local indicators

*Sub-Challenge: Information outside the System's View*

So far, we discussed that for standard components of IIR systems that support ISSs such as querying, receiving results, evaluating documents, saving documents we can measure the support towards the ISS by taking the user interactions and signals of positive success towards the ISS into account. This is a rather simplified model that assumes that all functionality the user needs is already available in the system. But for a lot of ISS this is simply not the case. Be it a simple filter functionality which is needed for the user's (more complex) task or a meaningful component for evaluating information side by side. If a needed functionality is not part of the system we cannot integrate it into an overall model and cannot compute the negative impact it may have on the overall usefulness. Here again knowledge about the overall goal and sub goals are needed. If we do not have this knowledge we cannot know about the user's need for this functionality. For this, only user-oriented studies can be helpful. Another point are user's cognitive processes such as comparing of information or learning that only happened in the user's mind and we do not know of. This can again only be made visible and understandable in explicit user studies and with techniques like think-aloud. Thus, the challenging issue is to specify user study designs that complement log-based usefulness analyses on a general level.

## 4. CONCLUSION

The usefulness model is an evaluation model for IIR which brings together the user's perspective with her/his goal and the system's perspective by supporting this goal. However, yet there is only initial work how this model can be operationalized. In this paper we discussed the challenges for the operationalization based on interaction data captured in log files. This can be a compromise between costly user studies and very specific TREC-style studies. We found that a key challenge is to find the task and overall goal. Much more research could be conducted here to better understand the user's intention. A good starting point could be for example research on standardized sets of information needs, information problems, task, goals and seeking strategies on the level of specific systems, situations, problems, contexts and domains. This would give the chance to categorize a user to a group with similar information needs and better support her/him with specific services. Capturing interaction data is the center of log-based methods, however, the interpretation of this data is complex. We discussed the challenging issues to be addressed on the different levels of system support.. From this it can be concluded that much more research is needed on capturing the task of the user at the one hand side, and on capturing positive signals for accomplishing a particular information seeking strategy on the other hand side.

**ACKNOWLEDGMENTS**

This work was partly funded by DFG, grant no. MA 3964/5-1; the AMUR project at GESIS. The authors thank the research group IIR at GESIS for fruitful discussions and suggestions.